\documentclass[twocolumn,aps]{revtex4}
\usepackage{graphicx}
\usepackage{dcolumn}
\usepackage{bm}

\begin{document}

\begin{titlepage}

\title{Towards Graphene Nanoribbon-based Electronics}

\author{Bing Huang, Qimin Yan, Zuanyi Li, and Wenhui Duan\footnote
{E-mail address: dwh@phys.tsinghua.edu.cn}}
\address{Department of Physics, Tsinghua University, Beijing
100084, People's Republic of China}
\date{\today}

\begin{abstract}

The successful fabrication of single layer graphene has greatly
stimulated the progress of the research on graphene. In this
article, focusing on the basic electronic and transport properties
of graphene nanoribbons (GNRs), we review the recent progress of
experimental fabrication of GNRs, and the theoretical and
experimental investigations of physical properties and device
applications of GNRs. We also briefly discuss the research efforts
on the spin polarization of GNRs in relation to the edge states.

\end{abstract}
%\keywords{graphene nanoribbon, electronic, transport, B, N, field
%effect transistor}

\maketitle

\draft

\vspace{2mm}

\end{titlepage}

\section{Introduction}

Graphene, one monolayer of carbon atoms tightly packed into a
two-dimensional honeycomb lattice, is actively being pursued as a
material for next-generation electronics due to its promising
electronic properties, such as high carrier mobility
\cite{Novoselov-S04, Berger-JPCB}, long phase coherence lengths
\cite{Berger}. On the other side, the unique two-dimensional atomic
structure of graphene implies unique confinement on electron system
and offers a perfect platform to explore the amazing physics
phenomenons, such as quantum Hall effect \cite{Gusynin,
Novoselov-S07, Zhang, Novoselov-N05} and massless Dirac fermions
\cite{Novoselov-N05, Zhou-NP06a, Bostwick, Andrei, Park-NP,
Li-NP08}.

The first task for experimentalists to study graphene electronics is
to fabricate high quality single layer graphene. Until now, several
different experimental methods have been proposed and realized to
prepare single layer (or few layers) graphene, including mechanical
exfoliation of highly oriented pyrolytic graphite
\cite{Novoselov-S04}, patterned epitaxially grown graphene on
silicon carbide or transition metal (e.g. Ru, Ni) substrates
\cite{Berger-JPCB, D. Usachov, Sutter, D. Martoccia}, liquid-phase
exfoliation of graphite \cite{Hernandez, Fan, Tung}, substrate-free
gas-phase synthesis \cite{Dato}, and chemical vapor deposition
\cite{Reina, Kim-N09}. The success in fabricating single layer
graphene has stimulated the extensive research efforts (both
theoretical and experimental) in graphene related research area.

The ultimate goal of the use of graphene in next-generation
electronics is to realize all-graphene circuit with functional
devices built from graphene layers or graphene nanoribbons (GNRs)
\cite{Areshkin-07b, Qimin}. As the basic building blocks of such
circuit, the concept of electronic devices based on graphene have
been proposed theoretically and realized by experiments recently,
such as field effect transistors \cite{Qimin, Liang, Meric, Song},
\emph{p-n} junctions \cite{J. R. Williams, D. A. Abanin-pn, B.
oyilmaz, Gorbachev}, gas molecule sensor \cite{F. Schedin, Wehling,
J. T. Robinson}, and so on.

In this article, we will focus our discussion on the basic
electronic and transport properties of GNRs and their application to
electronic devices. In particular, the theoretical investigations of
GNRs physics and the technical aspects of GNR based electronic
devices will be reviewed in detail. For other topics on the recent
experimental and theoretical research efforts on graphene, please
refer to the reviews by Katsnelson \cite{Katsnelson-MT}, Geim
\emph{et al.} \cite{A. K. Geim}, Beenakker \cite{Beenakker}, and
Castro Neto \emph{et al.} \cite{A. H. Castro Neto},

\section{Experimental Fabrication of Graphene Nanoribbons}

The realization of graphene electronics relies on the ability to
modify the electronic properties of finite-size graphenes (for
example, from semiconducting to metallic) by varying their size,
shape, and edge orientation. Such unique property compared to
traditional semiconductor materials, such as silicon, would
ultimately enable the design and miniaturization of future
electronic circuit by patterned graphene. One of the most
important issues in patterned graphene fabrication is the control
of the nanoribbon width. In order to take advantage of quantum
confinement effects in graphene, the ribbon width should go down
to nanometer scale. To realize the patterning of graphene with
nano-scale width, several different techniques have been proposed
including standard e-beam lithography (Fig. 1a) \cite{Han, Chen},
microscope lithography (Fig. 1b) \cite{L. Tapaszto, Weng,
Giesbers}, chemical method (Fig. 1c) \cite{Dai-Science}, metallic
nanoparticle etching \cite{Datta1}, and e-beam irradiation of
ultrathin poly(methylmethacrylate) (PMMA) \cite{Duan}. As shown in
Fig. 1a, the scanning electron microscopy (SEM) image reveals the
graphene can be patterned by traditional e-beam lithography
technique into nanoribbons with various widths ranging from 20 to
500 nm \cite{Chen}. Figure 1b shows 10-nm-wide nanoribbon etched
via scanning tunnelling microscope (STM) lithography. By setting
the optimal lithographic parameters, it is possible to cut GNRs
with suitably regular edges, which constitutes a great advance
towards the reproducibility of GNR-based devices \cite{L.
Tapaszto}. Figure 1c shows atomic force microscopy images of
chemically derived GNRs with various widths ranging from 50 nm to
sub-10 nm. These GNRs have atomic-scale ultrasmooth edges
\cite{Dai-Science}.

\begin{figure} [tbp]
\centering
\includegraphics[width=0.48\textwidth]{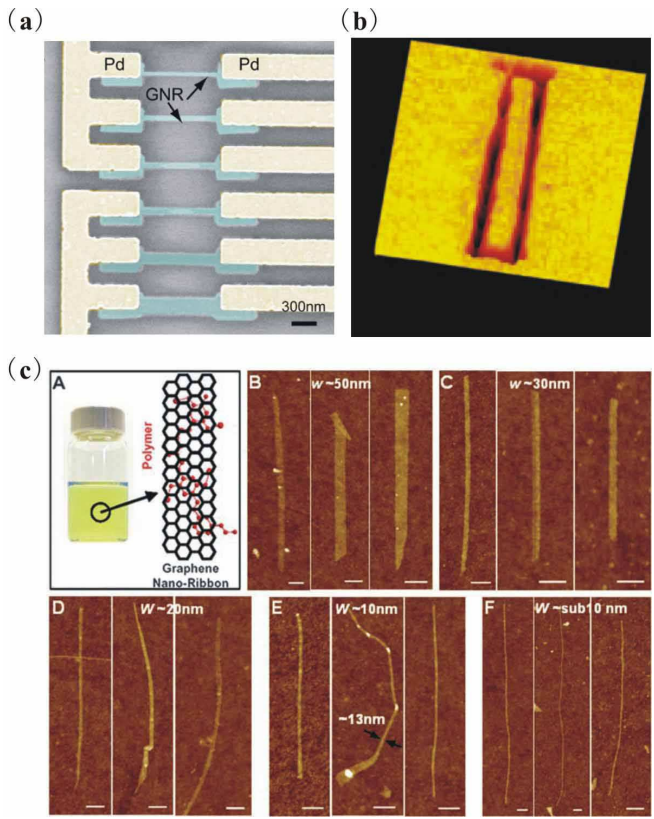}
\caption{\label{fig:fig1} (Color online) various GNRs got from
different experimental methods: (a) The SEM image of GNRs patterned
by e-beam lithography. Reprinted with permission from Ref.
\cite{Chen}, Z. Chen et al., Physica E \textbf{40}, 228 (2007).
\copyright~2007, Elsevier. (b) An 8-nm-wide 30$^\circ$ GNR bent
junction connecting an armchair and a zigzag nanoribbon etched by
STM lithography. Reprinted with permission from Ref. \cite{L.
Tapaszto}, L. Tapaszt\'{o} et al., Nat. Nanotechnol. \textbf{3}, 397
(2008). \copyright~2008, Nature Publishing Group. (c) GNRs are got
by using simple chemical methods. Reprinted with permission from
Ref. \cite{Dai-Science}, X. Li et al., Science \textbf{319}, 1229
(2008). \copyright~2008, American Association for the Advancement of
Science.}
\end{figure}

The electronic properties of GNRs exhibit a strong dependence on the
orientation of their edges. As two typical types, armchair GNRs
(AGNRs) and zigzag GNRs (ZGNRs) can be obtained by lithography
technology along the specific orientation on graphene (Fig. 2b)
\cite{Han, L. Tapaszto}. Actually, the detailed edge structures
(both armchair and zigzag) have already been clearly observed in
recent experiments \cite{Kobayashi, Y. Niimi, Zheng Liu}. One of the
most serious obstacle to graphene electronic application is the
reliable control of the edge structure of GNRs. Theoretical studies
predict that edge states (in a manner similar to the well-known
concept of surface states of a 3D crystal) in graphene are strongly
dependent on the edge termination and affect the physical properties
of GNRs \cite{Qimin, K. Nakada, K. Wakabayashi, Y. Miyamoto,
Son-PRL, Barone, Gunlycke-PRB08}. However, until now there is no
reliable experimental method which is able to exactly control the
edge structures and reduce their roughness. An interesting
experimental observation is that the band gaps of GNRs show little
orientation dependence \cite{Han} and all fabricated GNRs show
semiconducting behavior \cite{Dai-PRL}, which seems inconsistent
with theoretical results \cite{K. Nakada, K. Wakabayashi, Y.
Miyamoto}. One of the reason for such inconsistency comes from the
roughness of GNR edges and our explanation is also given in the
following part of the article. Another issue related to GNR edges is
the edge passivation. Since the dangling bonds from the edge carbon
atoms have relatively high chemical activity, there is the
possibility that other chemical elements present in the material
fabrication process (such as C, O, N, H and other chemical groups
formed by these atoms) would interact with the edge atoms and modify
the electronic properties of GNRs. To the best of our knowledge,
this issue has not been properly solved experimentally.

\begin{figure} [tbp]
\centering
\includegraphics[width=0.48\textwidth]{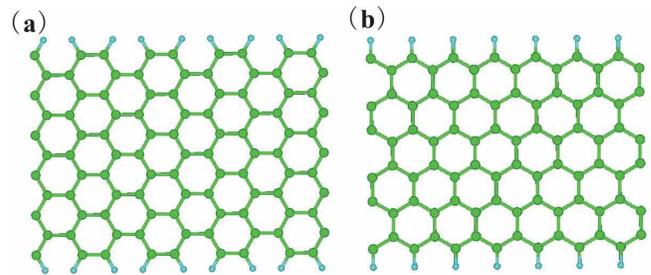}
\caption{\label{fig:fig2} (Color online) The structures of
H-passivated 11-AGNR (a) and 6-ZGNR (b), where big green balls and
small blue balls represent carbon atoms and hydrogen atoms,
respectively. Integer $N$ is their width index.}
\end{figure}

\section{Elementary Electronic and Transport Properties of Graphene Nanoribbons}

Next we will review some basic electronic and transport properties
of GNRs from the theoretical viewpoint. Figs. 2a and 2b show two
typical models of armchair and zigzag GNRs in first-principles or
other atomic-level electronic structure calculations, noting as
11-AGNR and 6-ZGNR, respectively. Here the numbers 11 and 6 are
defined as the width index, $N$. In order to remove the effect of
dangling bonds, the edges of GNRs are saturated by hydrogen atoms.
As geometrically terminated graphene, the electronic structure of
GNRs can be modelled by imposing appropriate boundary conditions
on Schr\"{o}dinger's equation with simple tight-binding (TB)
approximations based on $\pi$-states of carbon \cite{K. Nakada, K.
Wakabayashi}. Another way to get the band structure is to solve
two-dimensional Dirac's equation of massless free particles with
an effective speed of light to model GNR system \cite{L. Brey}.
Within these models, it is predicted that GNRs with
armchair-shaped edges can be either metallic or semiconducting
depending on their widths, as shown in Fig. 3a. On the other side,
the GNRs with zigzag-shaped edges are metallic with peculiar edge
states on both sides of ribbons regardless of their widths, as
shown in Fig. 4a. \cite{K. Nakada, K. Wakabayashi, Son-Nature}

\begin{figure} [tbp]
\centering
\includegraphics[width=0.48\textwidth]{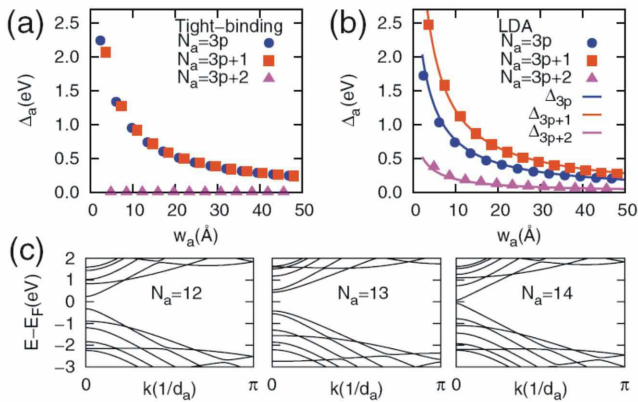}
\caption{\label{fig:fig3} (Color online) The variation of band gaps
of N$_{a}$-AGNRs as a function of width (w$_{a}$) obtained (a) from
TB calculations and (b) from first-principles calculations
(symbols). (c) First-principles band structures of N$_{a}$-AGNRs
with N$_{a}$= 12, 13, and 14, respectively.  Reprinted with
permission from Ref. \cite{Son-PRL}, Y. -W. Son et al., Phys. Rev.
Lett. \textbf{97}, 216803 (2006). \copyright~2006, American Physical
Society.}
\end{figure}

Further detailed \emph{ab initio} and GW quasiparticle
calculations show that all of the AGNRs exhibit semiconducting
behavior and the energy gaps decrease as a function of increasing
ribbon widths. The variation in energy gaps can be separated into
three distinct family behaviors \cite{Son-PRL, Barone, Qimin,
YangLi-GW}, as shown in Fig. 3b. As mentioned above, such
dependence of band gap on the geometrical structure of GNR offers
unique possibility to modify the electronic properties of GNRs
simply by controlling the width and edge orientation in order to
realize all-graphene functional devices.

\begin{figure} [tbp]
\centering
\includegraphics[width=0.48\textwidth]{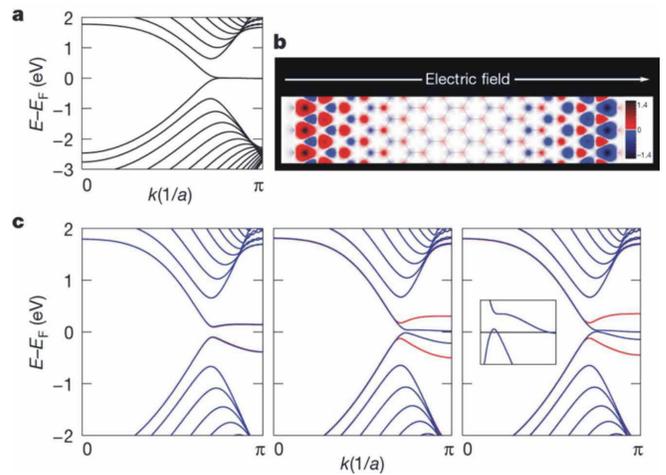}
\caption{\label{fig:fig4} (Color online) Electronic structures of
graphene nanoribbons. In all figures, the Fermi energy (E$_{F}$) is
set to zero. a, The spin-unpolarized band structure of a 16-ZGNR. b,
The spatial distribution of the charge difference between
$\alpha$-spin and $\beta$-spin for the ground state when there is no
external field. The magnetization per edge atom for each spin on
each sublattice is 0.43$\mu_B$ with opposite orientation, where
$\mu_B$ is the Bohr magneton. The graph is the electron density
integrated in the z direction, and the scale bar is in units of
10$^{-2}$e\AA$^{-2}$. c, From left to right, the spin-resolved band
structures of a 16-ZGNR with the external field of 0.0, 0.05 and
0.1V\AA, respectively. The red and blue lines denote bands of
$\alpha$-spin and $\beta$-spin states, respectively. Reprinted with
permission from Ref. \cite{Son-Nature}, Y. -W. Son et al., Nature
\textbf{444}, 347 (2006). \copyright~2006, Nature Publishing Group.}
\end{figure}

Upon inclusion of the spin degrees of freedom within density
functional theory (DFT) calculations, ZGNRs are predicted to have a
magnetic insulating ground state with ferromagnetic ordering at each
zigzag edge and antiparallel spin orientation between the two edges
\cite{Son-PRL, Son-Nature}, as shown in Fig. 4b. The spin
polarization originates from the edge states that introduce a high
density of state (DOS) at the Fermi energy. It can be qualitatively
understood in terms of the stoner magnetism of $sp$ electrons (in
analogy to conventional $d$ electrons), which occupy a very narrow
edge band and render instability of spin-band splitting \cite{Bing}.
What is more interesting, the zigzag GNRs show half-metallic
behavior when external transverse electric field is applied across
the ZGNRs along the lateral direction \cite{Son-Nature}, as shown in
Fig. 4c. However, such spin related half-metallic phenomenon becomes
weak with increasing ribbon width (since the total energy difference
per edge atom between spin-unpolarized and spin-polarized edge
states is only about 20 meV in their simulation system and decreases
with increasing width) and is not energetically stable if the width
of GNR is significantly larger than the decay length of the
spin-polarized edge states \cite{Kan-APL, Rudberg}. On the other
hand, it is predicted that the half-metallicity can be also achieved
in edge-modified or doped ZGNRs \cite{Hod, Kan-JACS, Sodi, Dutta}

Another important issue regarding the basic electronic structures of
GNRs relies on the edge states. Due to the presence of the edge
states, the $\pi$ and $\pi^{\ast}$ subbands of metallic ZGNRs (in
the spin-unpolarized state) do not cross with each other at the
Fermi level to span the whole energy range like metallic armchair
carbon nanotubes (CNTs) (the left panel of Fig. 5a). This leads to
the fact that the transport property of ZGNRs under a low bias
voltage (or a small potential step) is only determined by the
transmission between $\pi$ and $\pi^{\ast}$ subbands (as shown in
Fig. 5a). With the presence of such unique band structures, ZGNRs
exhibit two distinct transport behaviors depending on the existence
of $\sigma$ mirror symmetry with respect to the midplane between two
edges \cite{Zuanyi, Cresti, Akhmerov, Nakabayashi}, although all the
ZGNRs have similar metallic energy band structures. Since the $\pi$
and $\pi^{\ast}$ subbands of symmetric ZGNRs (i.e., width index $N$
is an even number) have opposite definite $\sigma$ parities, the
transmission between them is forbidden (the left panel of Fig. 5b).
For asymmetric ZGNRs (i.e., width index $N$ is an odd number),
however, their $\pi$ and $\pi^{\ast}$ subbands do not have definite
$\sigma$ parities, so the coupling between them can contribute to
about one conductance quantum (the right panel of Fig. 5b). This
transport difference can be clearly reflected in the
current-bias-voltage ($I$-$V_{\rm bias}$) characteristics of ZGNRs
by using the first-principle transport simulation, as shown in Fig.
5c. Although metallic armchair carbon nanotubes also have $\pi$ and
$\pi^{\ast}$ subbands with definite parities, such
symmetry-depending (or band-selective) $I$-$V_{\rm bias}$
characteristics cannot be observed in them because of the crossover
of their subbands, i.e., the absence of edge states. Recently,
theoretical work predicts a very large magnetoresistance in a
graphene nanoribbon device due to the existence of edge states
\cite{Woo}.

\begin{figure} [tbp]
\centering
\includegraphics[width=0.47\textwidth]{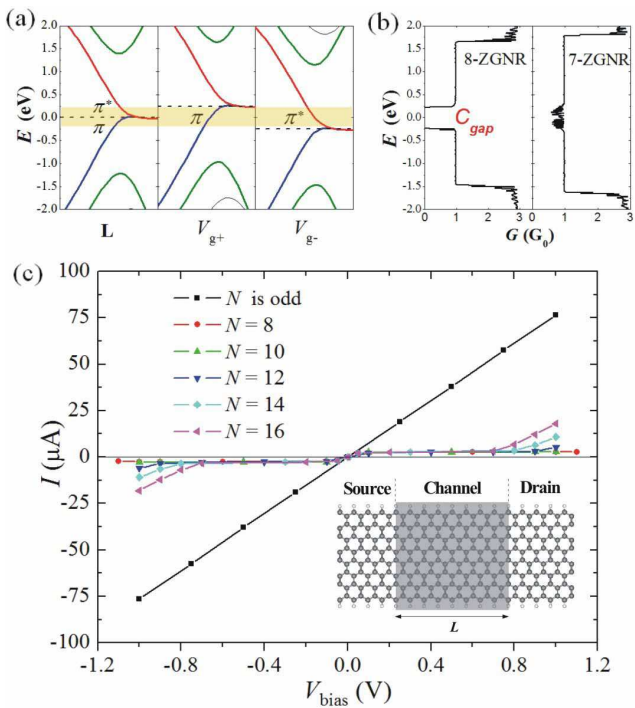}
\caption{\label{fig:fig5} (Color online) (a) Schematic band
structure around the Fermi level of a ZGNR under a positive
($V_{g+}$) and a negative ($V_{g-}$) potential step. (b) Conductance
of 8-ZGNR and 7-ZGNR under two potential steps shown in (a). (c)
$I$-$V_{\rm bias}$ curves of the two-probe system (see the inset)
made of ZGNRs with different widths $N$. Reprinted with permission
from Ref. \cite{Zuanyi}, Z. Li et al., Phys. Rev. Lett \textbf{100},
206802 (2008). \copyright~2008, American Physical Society.}
\end{figure}

Besides the fabrication and theoretical study of monolayer graphene
and GNRs, recent experimental \cite{Novoselov-bilayer, E. Rotenberg,
E. V. Castro} and theoretical \cite{M. I. Katsnelson, Kai-Tak Lama,
Eduardo V. Castro, Sahu-08} studies are also carried out on bilayer
graphene and GNRs. Theoretically, it is shown that the bilayer GNRs
and monolayer GNRs have some similar electronic properties such as
edge states localized at the zigzag edges and semiconducting
behavior of armchair bilayer GNRs \cite{Kai-Tak Lama, Eduardo V.
Castro, Sahu-08}. Experimentally, it is found that the bilayer
graphene has unique features such as anomalous integer quantum Hall
effects \cite{Novoselov-bilayer}, which is absent in single layer
graphene. And the size of energy gap of such bilayer structures can
be controlled by adjusting carrier concentration \cite{E. Rotenberg}
as well as by an external electric field \cite{E. V. Castro}. These
unique properties open an opportunity to implement bilayer graphene
or GNRs in various electronic applications.

\section{Edge disorder in Graphene Nanoribbons}

As mentioned above, current experimental techniques (such as
lithography) are not able to realize exact control of the edge
structures of GNRs and the edges are always very rough due to the
limitation of the fabrication technology \cite{Han, Chen}. There are
theoretical evidences that such edge disorders can significantly
change the electronic properties of GNRs \cite{Areshkin, Gunlycke,
Querlioz, TCLi, Lherbier, Cresti-NaR}, and lead to some unexpected
physics effect, such as the Anderson localization \cite{M.
Evaldsson, E. R. Mucciolo} and Coulomb blockade effect \cite{F.
Sols}. These effects have already been observed in lithographically
obtained graphene nanoribbons \cite{B. oyilmaz, Han, Stampfer-APL,
Stampfer-PRL, Kathryn Todd}.

The edge roughness is also crucial for the spin polarized
properties of GNRs. As we know, the magnetic properties of GNRs
depend on the highly degenerate edge states. In principle a
perfect edge structure is necessary for stabilizing magnetic
properties of GNRs as theoretically predicted. An important
question is, how robust the spin-polarized state is in the
presence of edge defects and impurities? The answer to this
question is not only scientifically interesting to better
understand the physical mechanism of spin polarization in GNRs but
also has important technological implications in the reliability
of GNRs as a new class of spintronic materials. First-principles
theoretical studies reveals the effect of edge vacancies and
substitutionally doped boron atoms, as typical examples of
structural edge defects and impurities, on the spin-polarization
of ZGNRs \cite{Bing}. The calculated energy difference between the
magnetic state [both antiferromagnetic (AF) and ferromagnetic
(FM)] and the nonmagnetic state is found to rapidly decrease with
increasing defect concentration and eventually decrease to zero
(nonmagnetic), as shown in Fig. 6. The critical defect (impurity)
concentration is found to be $\sim$ 0.10/\AA~ when the ribbon
width is larger than 2 nm.

\begin{figure} [tbp]
\centering
\includegraphics[width=0.4\textwidth]{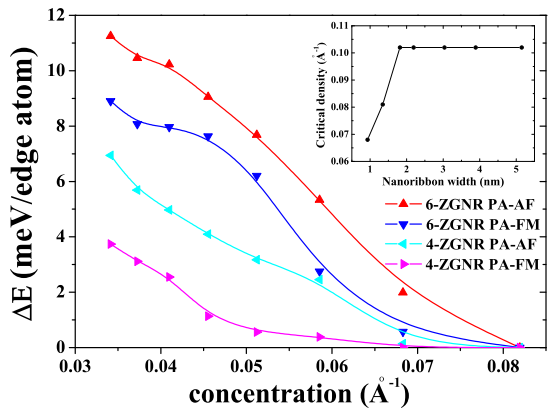}
\caption{\label{fig:fig6} (Color online) The energy difference per
edge atom between the magnetic (AF or FM) and paramagnetic (PA)
state as a function of vacancy concentration in the edge. Two
different ribbon widths of $N$=6 and $N$=4 are shown. The inset
shows the critical concentration as a function of the ribbon width
up to 5 nm. Reprinted with permission from Ref. \cite{Bing}, B.
Huang et al., Phys. Rev. B \textbf{77}, 153411 (2008).
\copyright~2008, American Physical Society.}
\end{figure}

Evidently, the magnetism in GNRs depends on a high density of
state (DOS) around the Fermi energy coming from the highly
degenerate edge states in a perfect ribbon edge ($E_{F}$) that
renders instability of spin polarization \cite{Bing}. The presence
of edge vacancies and impurities would decrease the DOS at $E_{F}$
since they do not contribute to the same edge state. From Stoner
model, such decrease of DOS will suppress the spin polarization of
GNR systems. Therefore, the practical realization of the spin
polarization in GNRs for spintronics applications could be rather
challenging \cite{Bing}. Recently, an interesting theoretical work
systematically studied the spin current in rough GNRs and
predicted that only GNRs with imperfect edges exhibit a nonzero
spin conductance while there is no spin current in perfect GNRs
\cite{Wimmer}. It confirms that the edge effect is of great
importance to spin related properties of GNRs.

Furthermore, the problem of edge passivation has not yet clearly
resolved by experiment until now. From the theoretical viewpoint,
the edge passivation can be well modeled by the modifications of
the hopping energies in the tight-binding approach \cite{Novikov}
or via additional phases in the boundary conditions \cite{Kane}.
Recent theoretical modeling and calculations have indicated that
the edge passivation has a strong effect on the electronic and
spin-polarized properties of GNRs \cite{Hod, Kan-JACS, Sodi}. The
possible passivation species include hydrogen, carbon, oxygen,
nitrogen, and other chemical groups. Further experimental works
are needed to explore the realistic edge structures of GNRs at
atomic scale and determine which types of edge passivation are
favorable.

\section{Transistors based on Graphene Nanoribbons}

The interesting and unique electronic properties of GNRs, such as
orientation and width dependence of transport behavior, offer great
possibilities for their electronic device applications. Compared
with other electronic materials, one of the most promising advantage
of GNRs is that GNR-based devices and even integrated circuits can
be fabricated by a single process of patterning a graphene sheet
\cite{Qimin}. Figures 7a-7c illustrate three basic device building
blocks: (i) metal-semiconductor junction, (ii) \emph{p-n} junction,
and (iii) hetero-junction, which can be, respectively, made by
patterned GNRs (i) along different direction, (ii) with different
edge doping, and (iii) with different widths. It was proposed that a
variety of devices can be constructed from these building blocks.
For example, a field effect transistor (FET) can be made simply by
two metal-semiconductor junctions, as shown in Fig. 7d. There are
some potential key advantages in designing and constructing device
architectures based on GNRs. The first advantage is the perfect
atomic interface, a feature that is difficult to achieve for the
interconnection between nanotubes of different diameter and
chirality. Second, it is generally difficult to find a robust method
to make contact with the molecular device unit, because there exists
usually a large contact resistance between the metal electrodes and
molecules (\emph{e.g.}, single-walled CNT) due to a very small
contact area. This difficulty may be circumvented by using GNRs,
because the GNR-based devices can be connected to the outside
circuits exclusively via metallic GNRs (or graphene), as illustrated
in Fig. 7d, which serve as extensions of metal electrodes to make
contact with the semiconducting GNRs so that an atomically smooth
metal-semiconductor interface is maintained with minimum contact
resistance. Last but not the least, the edges of GNRs may serve as
effective sites for doping. In principle, by introducing different
types of dopants at different sections of GNR edges, one can realize
a \emph{p-n} junction by selective doping, as shown in Fig. 7b.

\begin{figure} [tbp]
\centering
\includegraphics[width=0.38\textwidth]{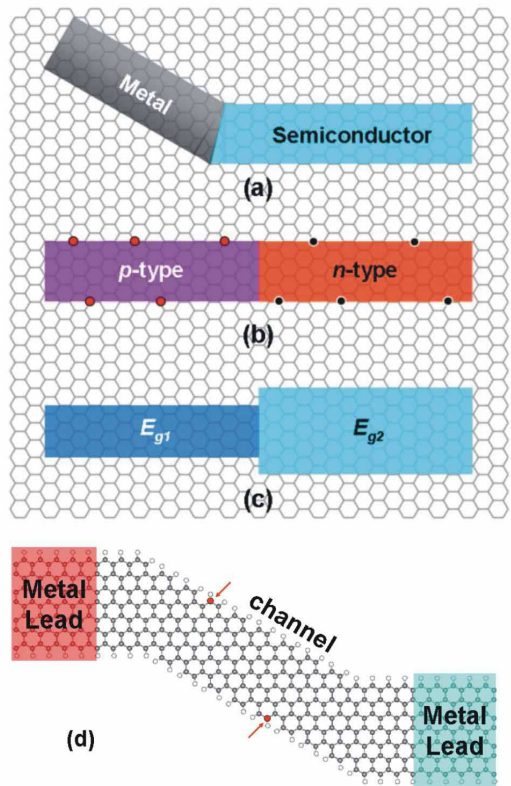}
\caption{\label{fig:fig7} (Color online) Schematics of three device
building blocks: (a) a metal-semiconductor junction between a zigzag
and an armchair GNR, (b) a \emph{p-n} junction between two armchair
GNRs with different edge doping, and (c) a heterojunction between
two armchair GNRs of different widths (band gaps). (d) Schematics of
a GNR-FET, made from one semiconducting 10-AGNR channel and two
metallic 7-ZGNR leads connected to two external metal electrodes.
Reprinted with permission from Ref. \cite{Qimin}, Q. Yan et al.,
Nano Lett. \textbf{7}, 1469 (2007). \copyright~2007, American
Chemical Society.}
\end{figure}

One of the most important electronic applications based on GNRs is
field effect transistors. Recently, experimental studies
\cite{Han, Chen, Dai-Science, Dai-PRL} have indicated the
possibility of fabricating GNR-based transistors. The advantage of
GNRs as an alternative material for transistors is that it could
bypass the chirality challenge of CNTs while retaining the
excellent electronic properties of graphene sheets, such as the
high $I_{\rm on}$/$I_{\rm off}$ ratio and excellent electron/hole
mobilities. The performance of one sub-10-nm GNR-FET in the latest
work from Dai's group is shown in Figs. 8a and 8b (the transfer
and output characteristics, respectively, for the GNR device with
the width of $\sim$ 2 $\pm$ 0.5 nm and the channel length of
$\sim$ 236 nm)\cite{Dai-PRL}. This device delivered $I_{\rm on}$
$\sim$ 4 $\mu$A at $V_{\rm ds}$ = 1 V, $I_{\rm on}$/$I_{\rm off}$
ratio $>$ 10$^{6}$ at $V_{\rm ds}$ = 0.5 V, subthreshold slope
$S$=d$V_{\rm gate}$/$\mathrm{d log}I$ $\sim$ 210 mV/decade and
transconductance $\sim$ 1.8 $\mu$S ($\sim$900 $\mu$S/$\mu$m). The
device performance is comparable with the best CNT-based
transistors. However, the Dirac point was not observed around zero
gate bias in this measurement, indicating \emph{p}-doping effects
at the edges or by physisorbed species during the chemical
treatment steps.

\begin{figure} [tbp]
\centering
\includegraphics[width=0.48\textwidth]{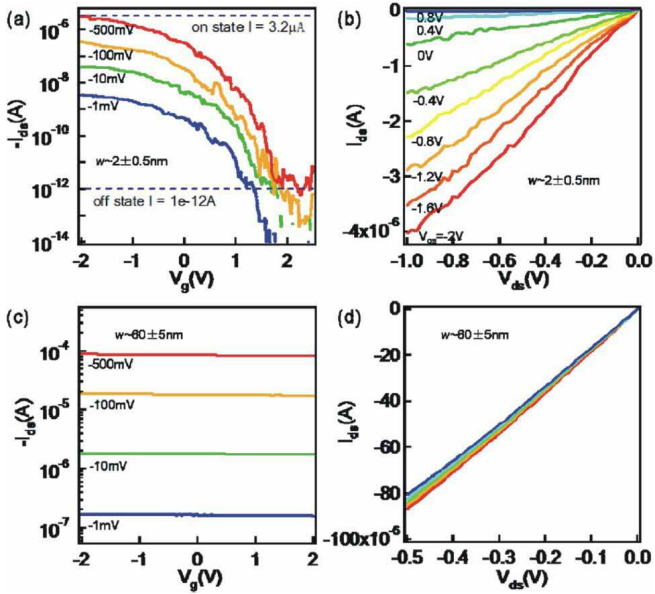}
\caption{\label{fig:fig8} (Color online) (a) and (b) Transistor
performance of GNR-FETs with width of $\sim$ 2 nm and channel length
of $\sim$ 236 nm [(c) and (d), width of $\sim$ 60 nm, and channel
length of $\sim$ 190 nm ]. (a) Transfer characteristics (current vs
gate voltage $I_{\rm ds}$-$V_{\rm gs}$) under various $V_{\rm ds}$.
$I_{\rm on}$/$I_{\rm off}$ ratio of $>$ 10$^{6}$ is achieved at room
temperature. (b) Output characteristics ($I_{\rm ds}$-$V_{\rm ds}$)
under various $V_{\rm gs}$. On current density is $\sim$ 2000
$\mu$A/$\mu$m in this device. (c) Transfer and (d) output
characteristics of the 60 nm width GNR-FET device. Reprinted with
permission from Ref. \cite{Dai-PRL}, X. Wang et al., Phys. Rev.
Lett. \textbf{100}, 206803 (2008). \copyright~2008, American
Physical Society.}
\end{figure}

Together with experimental progress on GNR-based transistors,
theoretical studies using semiclassical and quantum transport models
show that GNR-based FETs could have a similar performance as
CNT-based FETs and might outperform traditional Si-based FETs
\cite{Qimin, Ouyang-APL06, Liang-JAP, Ryzhii}. Figure 9 shows a
first-principles study on the performance of a typical GNR-based FET
made with a 5.91 nm long intrinsic semiconducting 10-AGNR channel
connected to two metallic 7-ZGNR leads (source and drain)
\cite{Qimin}. In Fig. 9a, the near-symmetric $I-V_{\rm gate}$ curve
shows an excellent ambipolar transistor with ON/OFF ratio $I_{\rm
on}$/$I_{\rm off}$ $\sim$ 2000 and subthreshold swing of $S$ $\sim$
60 mV/decade, which are comparable to those of high performance
CNT-FETs. Such the field effect can be clearly reflected in the
change of $I-V_{\rm bias}$ characteristics under different gate
voltages (Fig. 9b). Figure 9c shows the $I-V_{\rm gate}$ curves of
the GNR-FETs made from the same 10-AGNR channel with its length
ranging from 1.69 to 6.76 nm, from which the values of $S$ are
derived as a function of $L$ as shown in Figure 9d. Clearly, $S$
decreases with increasing $L$, and gradually approaches $\sim$60
mV/decade when $L$ becomes longer than 6 nm. Meanwhile, the
ON-current stays the same, independent of $L$, but the OFF-state
leakage current increases rapidly with decreasing $L$, which gives
rise to a large $S$. The performance of the ambipolar GNR-FETs made
of intrinsic semiconductor channels can be understood in terms of
metal-semiconductor tunneling junctionh within the semiclassical
band-bending model.

\begin{figure} [tbp]
\centering
\includegraphics[width=0.48\textwidth]{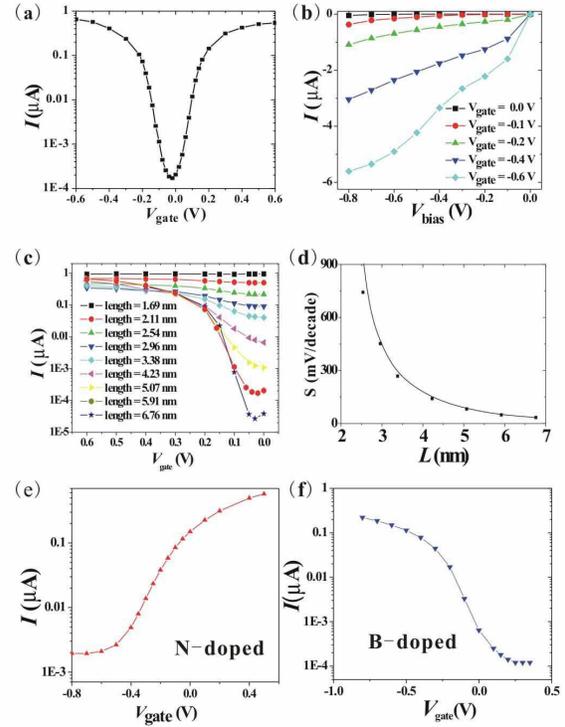}
\caption{\label{fig:fig9} (Color online) (a) $I-V_{\rm gate}$ curve
for a 5.91 nm long intrinsic 10-AGNR channel ($V_{\rm bias}$=20 mV).
(b) $I-V_{\rm bias}$ curves under different gate voltage ($V_{\rm
gate}$). (c) $I-V_{\rm gate}$ curves for different channel lengths
($V_{\rm bias}$=20 mV). (d) The subthreshold swing ($S$) as a
function of channel length $L$. (e),(f) $I-V_{\rm gate}$ curves for
a 5.91 nm long 10-AGNR channel with selective N and B doping,
respectively ($V_{\rm bias}$=20 mV). Reprinted with permission from
Ref. \cite{Qimin}, Q. Yan et al., Nano Lett. \textbf{7}, 1469
(2007). \copyright~2007, American Chemical Society.}
\end{figure}

Compared with the basic ambipolar FETs, it is well known that
\emph{n}-type (or \emph{p}-type) FETs serve as critical transistor
devices for digital electronics applications \cite{Avouris,
Dai-Nanolett}. To realize such device design based on GNRs, a
method was proposed using N (or B) atoms as selective dopants at
the channel region of perfect GNR-FETs (the positions of B or N
are indicated by arrows in Fig. 7d). Figure 9e (9f) shows the
calculated $I-V_{\rm gate}$ curves under $V_{\rm bias}$ = 20 mV,
exhibiting the typical behavior of a \emph{n}-type (\emph{p}-type)
FET\cite{Qimin}. It is suggested that all of the functional
transistor devices that work in traditional Si-based circuits
could be realized by GNRs and GNR-based junctions in principle.

\begin{figure} [tbp]
\centering
\includegraphics[width=0.48\textwidth]{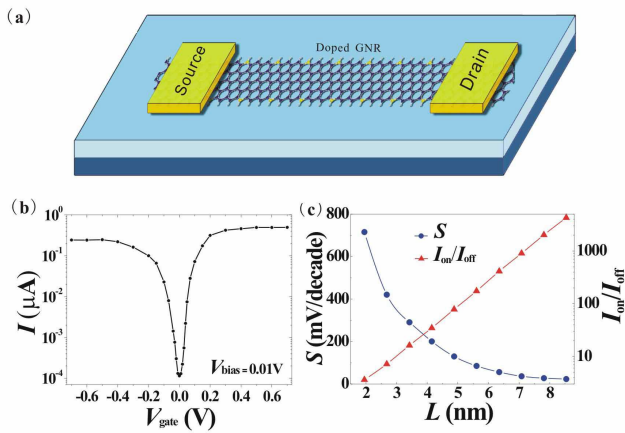}
\caption{\label{fig:fig10} (Color online) (a) The schematic
structure of the field effect transistor (FET) made from a single
5-ZGNR. The semiconducting channel is obtained by edge doping of N
in a finite-length region (the center region). (b) Simulated
$I$-$V_{\rm gate}$ curves of N-doped GNR-FETs under $V_{\rm bias}$ =
0.01 V. The channel length is 8.54 nm and the linear doping
concentration is 0.1365 \AA$^{-1}$. (c) The dependence of the
subthreshold swing $S$ (blue line) and the ON/OFF current ratio (red
line) on the channel length $L$. Reprinted with permission from Ref.
\cite{Bing-APL}, B. Huang et al., Appl. Phys. Lett. \textbf{91},
253122 (2007). \copyright~2007, American Institute of Physics.}
\end{figure}

Noting the current experimental difficulty to get an accurate
Z-shape junction (i.e., FET shown in Fig. 7d) due to the
limitation of lithography technique, a new type of field effect
transistor has also been proposed taking advantage of the
metal-semiconductor transition in ZGNRs induced by substitutional
doping of nitrogen or boron atoms at their edges \cite{Bing-APL},
as shown in Fig. 10a. Besides simplifying the fabrication process,
such a linear configuration can also increase the device density
in electronic circuits. Figure 10b shows a typical $I-V_{\rm
gate}$ curve for the N-doped GNR-FET (with the channel length of
8.54 nm) under the bias voltage $V_{\rm bias}$ = 0.01 V. Clearly,
the doped FET exhibits ambipolar characteristics, similar to the
Z-shape FETs. The relationship between the device performance and
the channel length is demonstrated by calculating $I-V_{\rm gate}$
curve of N-doped GNR-FETs as a function of the doped channel
length while keeping the bias voltage $V_{\rm bias}$ at 0.01 V. As
shown in Fig. 10c, the subthreshold swing $S$ of these doped
GNR-FETs decreases and the ON/OFF current ratio increases
exponentially. It can be seen that for good device performance
with small $S$ value (e.g., below 100 mV/decade) and high ON/OFF
current ratio (e.g., above 100), the doped channel length should
be longer than 5 nm. The minimum leakage current of those FETs
with the doped channels shorter than this critical length will be
greatly enhanced by direct tunneling, which lowers the device
performance.

Besides ideal case, some more practice issues concerning GNR-based
FETs are discussed in recent theoretical works. For example, the
effects of the various contact types and shapes on the performance
of Schottky-barrier-type GNR-FETs have been investigated
theoretically \cite{Liang-NaL}, which indicates that the
semi-infinite normal metal can potentially provide promising
performance. In addition, the effect of edge roughness and carrier
scattering on GNR-FETs have been studied \cite{Yoon, Basu, Ouyang}.
The presence of edge disorder significantly reduces ON-state
currents and increases OFF-state currents (the ON/OFF ratio
decreases), and introduces wide variability across devices. These
effects become weaker for GNRs with larger width and smoother edges.
However, the band gap decreases with increasing width, thereby
increasing the band-to-band tunneling mediated subthreshold leakage
current even with perfect GNRs. Obviously, without atomically
precise edge control during fabrication, it is hard to get reliable
and stable performance of GNR-FETs.

Due to their unusual basic properties, GNRs as well as graphene
are promising for a large number of applications \cite{A. K. Geim,
Cresti-NaR}, from spin filters \cite{Wimmer, Fazzio, Kan-JACS},
valley filters \cite{Rycerz}, to chemical sensors \cite{F.
Schedin, J. T. Robinson, Bing-JPCC}. GNRs can be chemically and/or
structurally modified in order to change its functionality and
hence its potential applications.

\section{Summary}

In summary, we review the basic electronic and transport
properties of graphene nanoribbons, and discuss recent theoretical
and experimental progress on GNR-based field effect transistors
from the viewpoint of device application. Due to the interesting
electronic and magnetic properties, GNRs have been demonstrated as
a promising candidate material for future post-silicon electronics
such as transport materials, field effect transistors, and spin
injection or filter. More experimental efforts will focus on
fabricating high quality nanoribbon samples with accurate control
of the edge structures.

\acknowledgments

This work was supported by the Ministry of Science and Technology
of China (Grant Nos. 2006CB605105 and 2006CB0L0601), and the
National Natural Science Foundation of China.

%\newpage


\begin{references}

\bibitem{Novoselov-S04} K. S. Novoselov, A. K. Geim, S. V. Morozov,
D. Jiang, Y. Zhang, S. V. Dubonos, I. V. Grigorieva, and A. A.
Firsov, Science {\bf 306}, 666 (2004).

\bibitem{Berger-JPCB} C. Berger, Z. Song, T. Li, X. Li, A. Y. Ogbazghi, R. Feng, Z. Dai,
A. N. Marchenkov, E. H. Conrad, P. N. First, and W. A. de Heer, J.
Phys. Chem. B {\bf 108}, 19912 (2004).

\bibitem{Berger} C. Berger, Z. Song, X. Li, X. Wu, N. Brown,
C. Naud, D. Mayou, T. Li, J. Hass, A. N. Marchenkov, E. H. Conrad,
P. N. First, and W. A. de Heer, Science {\bf 312}, 1191 (2006).

\bibitem{Gusynin} V. P. Gusynin, and S. G. Sharapov, Phys. Rev. Lett. {\bf 95}, 146801 (2005).

\bibitem{Novoselov-S07} K. S. Novoselov, Z. Jiang, Y. Zhang, S. V. Morozov,
H. L. Stormer, U. Zeitler, J. C. Maan, G. S. Boebinger, P. Kim, and
A. K. Geim, Science {\bf 315}, 1379 (2007).

\bibitem{Zhang} Y. Zhang, Y. W. Tan, H. L. Stormer, and P. Kim.
Nature (London) {\bf 438}, 201 (2005).

\bibitem{Novoselov-N05} K. S. Novoselov, A. K. Geim, S. V. Morozov, D. Jiang, M. I. Katsnelson,
I. V. Grigorieva, S. V. Dubonos, and A. A. Firsov, Nature (London)
{\bf 438}, 197 (2005).

\bibitem{Zhou-NP06a} S. Y. Zhou, G.-H. Gweon, J. Graf, A. V. Fedorov, C. D. Spataru,
R. D. Diehl, Y. Kopelevich, D.-H. Lee, S. G. Louie, and A. Lanzara,
Nature Phys. {\bf 2}, 595 (2006).

\bibitem{Bostwick} A. Bostwick, T. Ohta, T. Seyller, K. Horn, and E. Rotenberg,
Nature Phys. {\bf 3}, 36 (2007).

\bibitem{Andrei} G. Li and E. Y. Andrei, Nature Phys. {\bf 3}, 623 (2007).

\bibitem{Park-NP} C.-H. Park, L. Yang, Y.-W. Son, M. L. Cohen, and S. G. Louie,
Nature Phys. {\bf 4}, 213 (2008).

\bibitem{Li-NP08} Z. Q. Li, E. A. Henriksen, Z. Jiang, Z. Hao, M. C. Martin,
P. Kim, H. L. Stormer, and D. N. Basov, Nature Phys. {\bf 4}, 532
(2008).

\bibitem{D. Usachov} D. Usachov, A. M. Dobrotvorskii, A. Varykhalov, O. Rader, W. Gudat, A. M. Shikin,
and V. K. Adamchuk, Phys. Rev. B {\bf 78}, 085403 (2008).

\bibitem{Sutter} P. W. Sutter, J. Flege, and E. A. Sutter, Nature Mater.
{\bf 7}, 406 (2008).

\bibitem{D. Martoccia} D. Martoccia, P. R. Willmott, T. Brugger, M. Bjorck,
S. Gunther, C. M. Schleputz, A. Cervellino, S. A. Pauli, B. D.
Patterson, S. Marchini, J. Wintterlin, W. Moritz, and T. Greber,
Phys. Rev. Lett. {\bf 101}, 126102 (2008).

\bibitem{Hernandez} Y. Hernandez, V. Nicolosi, M. Lotya,F. M. Blighe,
Z. Sun, S. De, I. T. McGovern, B. Holland, M. Byrne, Y. K. Gunko, J.
J. Boland, P. Niraj, G. Duesberg, S. Krishnamurthy, R. Goodhue, J.
Hutchison, V. Scardaci, A. C. Ferrari, and J. N. Coleman, Nat.
Nanotechnol. {\bf 3}, 563 (2008).

\bibitem{Fan} X. Fan, W. Peng, Y. Li, X. Li, S. Wang, G. Zhang, and F. Zhang, Adv. Mater. {\bf 20}, 4490 (2008).

\bibitem{Tung} V. C. Tung, M. J. Allen, Y. Yang, and R. B. Kaner, Nat. Nanotechnol.
{\bf 4}, 25 (2009).

\bibitem{Dato} A. Dato, V. Radmilovic, Z. Lee, J. Phillips, and M. Frenklach, Nano Lett. {\bf 8}, 2012 (2008).

\bibitem{Reina} A. Reina, X. Jia, J. Ho, D. Nezich, H. Son, V. Bulovic, M. S. Dresselhaus, and J. Kong, Nano
Lett. {\bf 9}, 30 (2009).

\bibitem{Kim-N09} K. S. Kim, Y. Zhao, H. Jang, S. Y. Lee, J. M.
Kim, K. S. Kim, J.-H. Ahn, P. Kim, J.-Y. Choi, B. H. Hong, Nature
(London) {\bf 457}, 706 (2009).

\bibitem{Areshkin-07b} D. A. Areshkin, and C. T. White, Nano Lett. {\bf 7}, 3253 (2007).

\bibitem{Qimin} Q. Yan, B. Huang, J. Yu, F. Zheng, J. Zhang,
J. Wu, B.-L. Gu, F. Liu, and W. Duan, Nano Lett. {\bf 7}, 1469
(2007).

\bibitem{Liang} X. Liang, Z. Fu, and S. Y. Chou, Nano Lett. {\bf 7}, 3840 (2007).

\bibitem{Meric} I. Meric, M. Y. Han, A. F. Young, B. Ozyilmaz, P. Kim,
and K. L. Shepard, Nat. Nanotechnol. {\bf 3}, 654 (2008).

\bibitem{Song} X. Wu, M. Sprinkle, X. Li, F. Ming, C. Berger,
and Walt A. de Heer, Phys. Rev. Lett. {\bf 101}, 026801 (2008).

\bibitem{J. R. Williams} J. R. Williams, L. DiCarlo, and C. M. Marcus, Science {\bf 317}, 638 (2007).

\bibitem{D. A. Abanin-pn} D. A. Abanin and L. S. Levitov, Science {\bf 317}, 641 (2007).

\bibitem{B. oyilmaz} B. \"{O}yilmaz, P. Jarillo-Herrero, D. Efetov,
D. A. Abanin, L. S. Levitov, and P. Kim, Phys. Rev. Lett. {\bf 99},
166804 (2007).

\bibitem{Gorbachev} R. V. Gorbachev, A. S. Mayorov, A. K. Savchenko, D. W. Horsell, and F.
Guinea, Nano Lett. {\bf 8}, 1995 (2008).

\bibitem{F. Schedin} F. Schedin, A. K. Geim, S. V. Morozov, E. W. Hill, P. Blake,
M. I. Katsnelson, and K. S. Novoselov, Nature Mater. {\bf 6}, 652
(2007).

\bibitem{Wehling} T. O. Wehling, K. S. Novoselov, S. V. Morozov, E. E. Vdovin, M. I.
Katsnelson, A. K. Geim, and A. I. Lichtenstein, Nano Lett. {\bf 8},
173 (2008).

\bibitem{J. T. Robinson} J. T. Robinson, F. K. Perkins, E. S. Snow, Z. Wei, and P. E.
Sheehan, Nano Lett. {\bf 8}, 3137 (2008).

\bibitem{Katsnelson-MT} M. I. Katsnelson, Mater. Today {\bf 10}, 20 (2007).

\bibitem{A. K. Geim} A. K. Geim and K. S. Novoselov, Nature Mater. {\bf 6}, 183 (2007).

\bibitem{Beenakker} C. W. J. Beenakker, Rev. Mod. Phys. {\bf 80},
1337 (2008).

\bibitem{A. H. Castro Neto} A. H. Castro Neto, F. Guinea, N. M. R. Peres,
K. S. Novoselov, and A. K. Geim, Rev. Mod. Phys. {\bf 81}, 109
(2009).

\bibitem{Han} M. Y. Han, B. \"{O}yilmaz, Y. Zhang, and P. Kim, Phys. Rev. Lett. {\bf 98}, 206805 (2007).

\bibitem{Chen} Z. Chen, Y.-M. Lin, M. J. Rooks, and P. Avouris, Physica E {\bf 40}, 228 (2007).

\bibitem{L. Tapaszto} L. Tapaszt\'{o}, G. Dobrik, P. Lambin, and L. P. Bir\'{o}, Nat. Nanotechnol. {\bf 3}, 397 (2008).

\bibitem{Weng} L. Weng, L. Zhang, Y. P. Chen, and L. P. Rokhinson, Appl. Phys. Lett. {\bf 93}, 093107 (2008).

\bibitem{Giesbers} A. J. M. Giesbers, U. Zeitler, S. Neubeck, F. Freitag, K.S. Novoselov,
J. C. Maan, Solid St. Comm. {\bf 147}, 366 (2008).

\bibitem{Dai-Science} X. Li, X. Wang, L. Zhang, S. Lee and H. Dai, Science
{\bf 319}, 1229 (2008).

\bibitem{Datta1} S. S. Datta, D. R. Strachan, S. M. Khamis, and A. T. Charlie Johnson, Nano Lett. {\bf 8}, 1912 (2008).

\bibitem{Duan} H. Duan, E. Xie, L. Han, and Z. Xu, Adv. Mater. {\bf 20}, 3284 (2008).

\bibitem{Kobayashi} Y. Kobayashi,1, K. Fukui, T. Enoki, K. Kusakabe, and Y. Kaburagi, Phys. Rev. B {\bf 71}, 193406 (2005).

\bibitem{Y. Niimi} Y. Niimi, T. Matsui, H. Kambara, K. Tagami, M. Tsukada,
and H. Fukuyama, Phys. Rev. B {\bf 73}, 085421 (2006).

\bibitem{Zheng Liu} Z. Liu, K. Suenaga, Peter J. F. Harris, and S. Iijima,
Phys. Rev. Lett. {\bf 102}, 015501 (2009).

\bibitem{K. Nakada} K. Nakada, M. Fujita, G. Dresselhaus and M. S.
Dresselhaus, Phys. Rev. B {\bf 54}, 17954 (1996).

\bibitem{K. Wakabayashi} K. Wakabayashi, M. Fujita, H. Ajiki, and M. Sigris,
Phys. Rev. B {\bf 59}, 8271 (1999).

\bibitem{Y. Miyamoto} Y. Miyamoto, K. Nakada, and M. Fujita, Phys. Rev.
B {\bf 59}, 9858 (1999).

\bibitem{Son-PRL} Y. -W. Son, M. L. Cohen, and S. G. Louie, Phys. Rev. Lett. {\bf 97}, 216803 (2006).

\bibitem{Barone} V. Barone, O. Hod, and G. E. Scuseria, Nano Lett. {\bf 6}, 2748 (2006).

\bibitem{Gunlycke-PRB08} D. Gunlycke and C. T. White, Phys. Rev.
B {\bf 77}, 115116 (2008).

\bibitem{Dai-PRL} X. Wang, Y. Ouyang, X. Li, H. Wang,
and H. Dai, Phys. Rev. Lett. {\bf 100}, 206803 (2008).

\bibitem{L. Brey} L. Brey, and H. A. Fertig, Phys. Rev. B {\bf 73}, 235411 (2006).

\bibitem{Son-Nature} Y.-W. Son, M. L. Cohen, and S. G. Louie, Nature (London) {\bf 444}, 347 (2006).

\bibitem{YangLi-GW} L. Yang, C.-H. Park, Y.-W. Son, M. L. Cohen, and S. G. Louie, Phys. Rev. Lett. {\bf 99}, 186801 (2007).

\bibitem{Bing} B. Huang, F. Liu, J. Wu, B.-L. Gu, and W. Duan,
Phys. Rev. B {\bf 77}, 153411 (2008).

\bibitem{Kan-APL} E.-J. Kan, Z. Li, J. Yang, and J. G. Hou,
Appl. Phys. Lett. {\bf 91}, 243116 (2007).

\bibitem{Rudberg} E. Rudberg, P. Sa{\l}ek, and Y. Luo, Nano Lett. {\bf 7},
2211 (2007).

\bibitem{Hod} O. Hod, V. Barone, J. E. Peralta, and G. E. Scuseria, Nano Lett. {\bf 7}, 2295
(2007).

\bibitem{Kan-JACS} E.-J. Kan, Z. Li, J. Yang, and J. G. Hou, J. Am. Chem. Soc. {\bf 130}, 4224
(2008).

\bibitem{Sodi} F. Cervantes-Sodi, G. Cs\'{a}nyi, S. Piscanec, and A. C. Ferrari, Phys. Rev. B {\bf 77}, 165427 (2008).

\bibitem{Dutta} S. Dutta and S. K. Pati, J. Phys. Chem. B {\bf 112}, 1333 (2008).

\bibitem{Zuanyi} Z. Li, H. Qian, J. Wu, B.-L. Gu, and W. Duan, Phys. Rev. Lett. {\bf 100}, 206802 (2008).

\bibitem{Akhmerov} A. R. Akhmerov, J. H. Bardarson, A. Rycerz, and C. W. J. Beenakker, Phys. Rev. B {\bf 77}, 205416 (2008).

\bibitem{Cresti} A. Cresti, G. Grosso, and G. Pastori Parravicini, Phys. Rev. B {\bf 77}, 233402 (2008).

\bibitem{Nakabayashi} J. Nakabayashi, D. Yamamoto, and S. Kurihara, Phys. Rev. Lett. {\bf 102}, 066803 (2009).

\bibitem{Woo} W. Y. Kim, and K. S. Kim, Nat. Nanotechnol. {\bf 3}, 408 (2008).

\bibitem{Novoselov-bilayer} K. S. Novoselov, E. McCann, S. V. Morozov, V. I. Falko, M. I. Katsnelson,
U. Zeitler, D. Jiang, F. Schedin, and A. K. Geim, Nat. Phys. {\bf
2}, 177 (2006).

\bibitem{E. Rotenberg} T. Ohta, A. Bostwick, T. Seyller, K. Horn, and E. Rotenberg,  Science {\bf 313}, 951 (2006).

\bibitem{E. V. Castro} E. V. Castro, K. S. Novoselov, S. V. Morozov, N. M. R. Peres, J. M. B.
Lopes dos Santos, J. Nilsson, F. Guinea, A. K. Geim, and A. H.
Castro Neto, Phys. Rev. Lett. {\bf 99}, 216802 (2007).

\bibitem{M. I. Katsnelson} M. I. Katsnelson, K. S. Novoselov, and A. K. Geim, Nat. Phys. {\bf 2}, 620 (2006).

\bibitem{Kai-Tak Lama} K.-T. Lam and G. Liang, Appl. Phys. Lett. {\bf 92}, 223106 (2008).

\bibitem{Eduardo V. Castro} E. V. Castro, N. M. R. Peres, J. M. B. Lopes dos Santos, A. H. Castro Neto,
and F. Guinea, Phys. Rev. Lett. {\bf 100}, 026802 (2008).

\bibitem{Sahu-08} B. Sahu, H. Min, A. H. MacDonald, and S. K. Banerjee, Phys. Rev. B {\bf 78}, 045404 (2008).

\bibitem{Areshkin} D. A. Areshkin, D. Gunlycke, and C. T. White, Nano Lett. {\bf 7}, 204 (2007).

\bibitem{Gunlycke} D. Gunlycke, D. A. Areshkin, and C. T. White, Appl. Phys. Lett. {\bf 90}, 12104 (2007).

\bibitem{Querlioz} D. Querlioz, Y. Apertet, A. Valentin, K. Huet, A. Bournel,
S. Galdin-Retailleau, and P. Dollfus, Appl. Phys. Lett. {\bf 92},
042108 (2008).

\bibitem{TCLi} T. C. Li, and S.-P. Lu, Phys. Rev. B {\bf 77}, 085408 (2008).

\bibitem{Lherbier} A. Lherbier, B. Biel, Y.-M. Niquet, and S. Roche, Phys. Rev. Lett. {\bf 100}, 036803 (2008).

\bibitem{Cresti-NaR} A. Cresti, N. Nemec, B. Biel, G. Niebler, F. Triozon, G. Cuniberti,
and S. Roche, Nano Res. {\bf 1}, 361 (2008).

\bibitem{M. Evaldsson} M. Evaldsson, I. V. Zozoulenko, H. Xu, and T. Heinzel,
Phys. Rev. B {\bf 78}, 161407 (2008).

\bibitem{E. R. Mucciolo} E. R. Mucciolo, A. H. Castro Neto, and C. H.
Lewenkopf, Phys. Rev. B {\bf 79}, 075407 (2009).

\bibitem{F. Sols} F. Sols, F. Guinea, and A. H. Castro Neto,
Phys. Rev. Lett. {\bf 99}, 166803 (2007).

\bibitem{Stampfer-APL} C. Stampfer, J. G\"{u}ttinger, F. Molitor, D. Graf, T. Ihn,
and K. Ensslin, Appl. Phys. Lett. {\bf 92}, 012102 (2008).

\bibitem{Stampfer-PRL} C. Stampfer, J. G\"{u}ttinger, S. Hellm\"{u}ller, F. Molitor,
K. Ensslin, and T. Ihn, Phys. Rev. Lett. {\bf 102}, 056403 (2009).

\bibitem{Kathryn Todd} K. Todd, H.-T. Chou, S. Amasha, and D. Goldhaber-Gordon, Nano Lett. {\bf 9},
416 (2009).

\bibitem{Wimmer} M. Wimmer, I. Adagideli, S. Berber, D. Tomanek,
and K. Richter, Phys. Rev. Lett. {\bf 100}, 177207 (2008).

\bibitem{Novikov} D. S. Novikov, Phys. Rev. Lett. {\bf 99}, 056802 (2007).

\bibitem{Kane} C. L. Kane, and E. J. Mele, Phys. Rev. Lett. {\bf 78}, 1932 (1997).

\bibitem{Ouyang-APL06} Y. Ouyang, Y. Yoon, J. K. Fodor, and J. Guo, Appl. Phys. Lett. {\bf 89}, 203107 (2006).

\bibitem{Liang-JAP} G. Liang, N. Neophytou, M. S. Lundstrom, and D. E. Nikonov, J. Appl. Phys. {\bf 102}, 054307
(2007).

\bibitem{Ryzhii} V. Ryzhii, M. Ryzhii, A. Satou, and T. Otsuji, J. Appl. Phys. {\bf 103}, 094510
(2008).

\bibitem{Avouris} P. Avouris, Acc. Chem. Res. {\bf 35}, 1026
(2002).

\bibitem{Dai-Nanolett} A. Javey, J. Guo, D. B. Farmer, Q. Wang, D. Wang, R. G. Gordon,
M. Lundstrom, and H. Dai, Nano Lett. {\bf 4}, 447 (2004).

\bibitem{Bing-APL} B. Huang, Q. Yan, G. Zhou, J. Wu, B.-L. Gu,
W. Duan, and F. Liu, Appl. Phys. Lett. {\bf 91}, 253122 (2007).

\bibitem{Liang-NaL} G. Liang, N. Neophytou, M. S. Lundstrom, and D. E. Nikonov, Nano Lett. {\bf 8}, 1819
(2008).

\bibitem{Yoon} Y. Yoon and J. Guo, Appl. Phys. Lett. {\bf 91}, 073103 (2007).

\bibitem{Basu} D. Basu, M. J. Gilbert, L. F. Register, S. K. Banerjee,
and A. H. MacDonald, Appl. Phys. Lett. {\bf 92}, 042114 (2008).

\bibitem{Ouyang} Y. Ouyang, X. Wang, H. Dai, and J. Guo, Appl. Phys. Lett. {\bf 92}, 243124 (2008).

\bibitem{Fazzio} T. B. Martins, R. H. Miwa, A. J. R. da Silva, and A. Fazzio, Phys. Rev. Lett. {\bf 98}, 196803 (2007).

\bibitem{Rycerz} A. Rycerz, J. Tworzyd\l o, and C. W. J. Beenakker, Nature Phys. {\bf 3}, 172 (2007).

\bibitem{Bing-JPCC} B. Huang, Z. Li, Z. Liu, G. Zhou, S. Hao, J. Wu, B.-L. Gu, and W. Duan, J. Phys. Chem. C {\bf 112}, 13442 (2008).


\end{references}
\end{document}